\documentstyle[aps,prb,preprint,epsfig]{revtex}

\renewcommand{\vec}[1]{\relax\ifmmode\mathchoice
{\mbox{\boldmath$\relax\displaystyle#1$}}
{\mbox{\boldmath$\relax\textstyle#1$}}
{\mbox{\boldmath$\relax\scriptstyle#1$}}
{\mbox{\boldmath$\relax\scriptscriptstyle#1$}}\else
\hbox{\boldmath$\relax\textstyle#1$}\fi}

\newcommand{\be}{\begin{equation}}
\newcommand{\ee}{\end{equation}}
\newcommand{\refkl}[1]{(\ref{#1})}
\renewcommand{\v}[1]{\mbox{\boldmath$#1$}}
\renewcommand{\sup}[1]{^{\rm #1}}
\newcommand{\sub}[1]{_{\rm #1}}
\newcommand{\rhomax}{\rho_{\rm max}}
\begin{document}

\centerline{\bf Numerical Simulation of Macroscopic Traffic
Equations}

\centerline{Dirk Helbing and Martin Treiber}
\footnotetext{\vskip -16pt \hskip -16pt \vbox{\parindent=0pt 
Dirk Helbing and Martin Treiber are members of the II. Institute
of Theoretical Physics, University of Stuttgart, Pfaffenwaldring
57, D-70550 Stuttgart, Germany.}}
\bigskip

The increasing need for efficient traffic optimization measures is
making reliable, fast, and robust methods for traffic simulation
more and more important. Apart from the development of cellular
automata models of traffic flow, %\cite{Nagel,Helb-nature} 
this need
has stimulated studies of suitable numerical algorithms for the
solution of macroscopic traffic equations based on partial
differential equations.
\cite{Dag95c,LePre92,MiBeLi,PaPiLy93,Lebacque,Ans,BuNeNa,BaAuToMa,Transp-KronKon}

The numerical integration of partial differential equations is a
particularly difficult task, and there is no generally
applicable method. In contrast to ordinary
differential equations, the most natural explicit finite difference
methods such as replacing differentials by forward,
backward, or central finite differences often are
numerically unstable,\cite{Anderson,NumRec} 
even in the limit of
very small discretizations in space and time. For example, a
necessary condition for stability is that the maximum speed at which
information about the solution can be propagated by the numerical
method must be always greater than the propagation velocity of the
exact solution.\cite{Anderson,Leveque} 

In general, the numerical
solution of partial differential equations requires special
methods, which work only under certain conditions.\cite{Anderson,Leveque}
Implicit integration methods are usually more
stable, but they require the frequent solution of
linear systems with multi-diagonal matrices. In this column, we
will discuss only explicit methods, because they are useful for the
varying boundary conditions found in realistic traffic simulations,
where data is continuously fed into the
simulation. In addition, explicit methods are more flexible
for the simulation of on- and off-ramps or entire road networks.

\smallskip \noindent {\bf Macroscopic Traffic Models}

Because the number of vehicles is conserved, all
macroscopic traffic models are based on the continuity equation for
the vehicle density $\rho(x,t)$ per lane at position $x$ and time
$t$:
\be 
 \frac{\partial \rho}{\partial t}
 + \frac{\partial (\rho V)}{\partial x} = \pm \nu^\pm (x,t) \, ,
\label{contin}
\ee
where $V(x,t)$ is the average vehicle velocity. According to
Eq.~(\ref{contin}), the temporal change 
$\partial \rho/\partial t$ of the vehicle density 
is given by the spatial change
$-\partial Q/\partial x$ of the traffic flow $Q(x,t) = \rho(x,t)
V(x,t)$ and the rate $\nu^\pm (x,t) \ge 0$
of vehicles entering (+) or leaving ($-$) the
highway at on- or off-ramps. 

To describe time and spatially varying velocities such as
occur in emergent traffic jams and stop-and-go traffic, we need a
dynamical velocity equation. For most continuous models, this
equation can be written as\cite{Helb-gas96}
\begin{equation}
 \frac{\partial V}{\partial t}
 + \underbrace{V \frac{\partial V}{\partial x}}_{\rm Transport\ Term}
 = \underbrace{- \frac{1}{\rho} \frac{\partial {P}}
 {\partial x}}_{\rm Pressure\ Term}
 + \underbrace{\frac{1}{\tau} ( V_{\rm e} - V )}_{\rm Relaxation\ Term} \! .
\label{geschwin}
\end{equation} 
According to Eq.~(\ref{geschwin}), the change 
$\partial V/\partial t$ of the average vehicle velocity is given by
three terms. The transport term originates from the propagation of
the velocity profile with the velocity $V$ of the vehicles. The
pressure term reflects either an anticipation of spatial changes
in the traffic situation or dispersion effects due to a finite
variance of the vehicle velocities. The relaxation term describes
the adaptation to a dynamic equilibrium velocity
$V_{\rm e}$ with a relaxation time
$\tau$.

All forms of congested traffic seem to have almost universal
properties which are largely independent of the initial conditions
and the spatially averaged density. For example, the characteristic
outflow $Q_{\rm out}$ from traffic jams is about
$1800\pm 200$ vehicles per kilometer and lane, and a typical
dissolution velocity $C$ of about $-15\pm 5$ kilometers per 
hour.\cite{Kerner-rehb96} This universality arises from the highly
correlated state of motion produced by traffic
congestions.\cite{Transp-Kerner} In particular, the
outflow $Q_{\rm out}$ is related to the time interval between successive
departures from the traffic jam.
Therefore, the outflow is almost independent of the kind and
density of congested traffic. As a consequence of the constant
outflow, the dissolution velocity of traffic jams is nearly
constant as well. These observations and the transition from free
to ``synchronized'' congested traffic\cite{Kerner-sync} are
correctly described by the nonlocal, gas-kinetic-based traffic
model,~\cite{GKT,Letter,GKT-scatter,Phase} which we now introduce. 

\smallskip \noindent {\bf The Nonlocal, Gas-Kinetic-Based Traffic
Model}

Our approach is to derive macroscopic traffic models from
gas-kinetic traffic 
equations,\cite{Helb-gas96,GKT} which are
obtained from ``microscopic'' models of driver-vehicle
behavior.\cite{GKT} Gas-kinetic traffic models have been
proposed
earlier, %\cite{Prigogine-traff,Paveri-Fontana,Phillips,Nelson} 
but the correct treatment of the most interesting regime of moderate
and high densities was still an open problem.

We have managed to evaluate the vehicular interaction term of the
gas-kinetic traffic model almost exactly.\cite{GKT}
The analytical result can be represented by the nonlocal, dynamical
equilibrium velocity $V_{\rm e}$: 
\be
\label{Vedyn}
 V_{\rm e} = V_0\left[
 1 - \frac{\theta+\theta'}{2 A(\rhomax)}
 \left( \frac{\rho' T}{1-\rho'/\rhomax} \right)^2
B(\delta_V) \right] \, .
\ee
According to Eq.~(\ref{Vedyn}), $V_{\rm e}$ is given by
the desired (maximum) velocity $V_0$, reduced by a term which
reflects necessary deceleration maneuvers. 
Here, $\rho_{\rm max}$ is the maximum vehicle density, and
$T$ is the average time headway at large densities. 
For example, German authorities require that the distance in
meters to the front vehicle be not less than half the velocity in km
per hour, which gives a time headway of $T=1.8\, {\rm s}$. For the
intra-lane variance $\theta$, we assume the constitutive relation
$\theta= A(\rho) V^2$ where $A(\rho)$ is given by
Eq.~\refkl{theta}. The prime indicates that the variable is
calculated at the advanced ``interaction point'' $x' = x +
\gamma (1/\rho_{\rm max} + VT)$ with $1 \le \gamma < 2$
rather than at the actual position $x$. This factor accounts for the
fact that drivers anticipate the behavior of vehicles in front of them.
The monotonically increasing ``Boltzmann
factor'' 
\begin{equation}
\label{B}
B(\delta_V) = 2 \left[ 
\delta_V \frac{e^{-\delta_V^2/2}}{\sqrt{2\pi}}
 + (1+\delta_V^2) 
\! \int_{-\infty}^{\delta_V} \! dy \,
\frac{e^{-y^2/2}}{\sqrt{2\pi}}\right]
\end{equation}
describes\cite{GKT} the dependence of the
braking interaction on the dimensionless velocity difference 
$\delta_V = (V-V')/\sqrt{\theta+\theta'}$
between the actual location $x$ 
and $x'$. In homogeneous traffic, we have $B(0)=1$.
If the preceding cars are much slower ($\delta_V \gg 0$),
the interaction strength given by $B$ is particularly large, and
it follows that
$B(\delta_V) = 2 \delta_V^2$. For
$\delta_V
\ll 0$, we have $B(\delta_V) \approx 0$. That is, because the
distance to the next vehicle is increasing, the vehicle will not brake, 
even if its 
distance is smaller than the safe distance.

Finally, the dynamics of
the intra-lane variance can be approximated by the constitutive
relation
\begin{equation}
\label{theta}
 \theta(x,t) = A(\rho)V^2(x,t) = \left[ A_0 + \Delta A \tanh 
\left(\frac{\rho(x,t)-\rho_{\rm c}}{\Delta\rho} \right)
\right] V^2(x,t) \, ,
\end{equation}
where the coefficients $A_0=0.008$, $\Delta A = 0.02$, $\rho_{\rm c}
= 0.27\rho_{\rm max}$, and $\Delta\rho = 0.05\rho_{\rm max}$
have been obtained from single-vehicle data on a section of the
Dutch motorway A9.\cite{GKT} Simulations of sections of other
motorways, for example, the German motorway A8, give a somewhat
lower value of $\Delta A=0.01$, which we will use in the
following. Equation~(\ref{theta}) shows that the standard deviation 
$\sqrt{\theta}$ of vehicle velocities is proportional to the average 
velocity $V$, with a density-dependent proportionality factor 
which is small for the density range found
in free traffic. 
The velocity variance also enters the gas-kinetic traffic
pressure
\begin{equation}
P(x,t) = \rho(x,t) \theta(x,t) 
\, .
\end{equation}
If not explicitly stated otherwise,
the simulation results presented here were calculated with
the model parameters 
$V_0 =110\, {\rm km/h}$, $\tau=32$ s, $T=1.8$ s, $\rhomax=160$
vehicles/km, and
$\gamma=1.2$. 

\begin{figure}
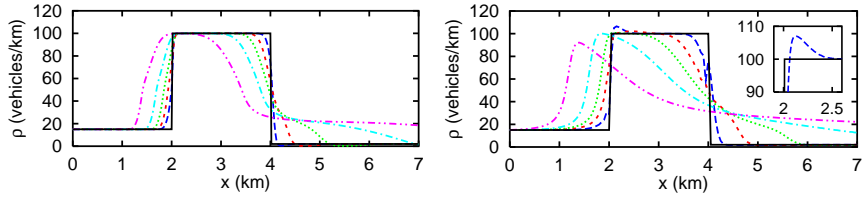

\begin{center} 
\begin{picture}(330,80)
\put(-10,0){\includegraphics[width=170\unitlength]{fig1a.cps}} 
\put(165,10){\includegraphics[width=158\unitlength]{fig1b.cps}}
\end{picture}
\end{center}
\caption{\label{figModel1}Simulation of discontinuous initial
conditions for (a) the non-local, gas-kinetic-based traffic model, 
and (b) for a typical traffic model with a viscosity term.\cite{lee}
The density profile is shown at $t=0\, {\rm s}$
(---), $10\,s$ (--~--), $30\,s$ (-~-~-), $60\, {\rm s}$ ($\cdots$),
$120\, {\rm s}$ (--~$\cdot$~--), and $240\, {\rm s}$
(--~$\cdot$~$\cdot$). The inset of (b) shows an unrealistic 
detail of the profile (upstream front) at $t=10\, {\rm s}$. Also note that, in
(b), the upstream front propagates too fast.}
\end{figure}

\begin{figure}
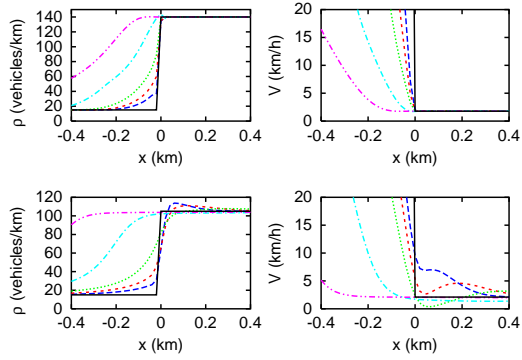

\begin{center}
\hspace*{-20\unitlength}
\includegraphics[width=100\unitlength]{fig2a.cps} 
\hspace*{-10\unitlength}
\includegraphics[width=100\unitlength]{fig2b.cps} \\
 \hspace*{-20\unitlength}
\includegraphics[width=100\unitlength]{fig2c.cps} 
\hspace*{-10\unitlength}
\includegraphics[width=100\unitlength]{fig2d.cps}
\end{center}
\caption{\label{figModel2}
First stages of the density and velocity profiles evolving from a
discontinuous upstream front (solid lines) corresponding, for
example, to an end of a traffic jam behind a curve. The
lines correspond to 
$t=0$\,s (---), 5\,s (--~--), 10\,s (-~-~-), 20\,s ($\cdots$),
60\,s (--~$\cdot$~--), and 120\,s (--~$\cdot$~$\cdot$).  
Parts (a) and (b) show
simulations with the non-local, gas-kinetic-based model ($\Delta
x=20$\,m,
$\Delta t=0.1$\,s), while (c) and (d) correspond to a typical
traffic model with a viscosity term.\cite{lee}
To obtain comparable equilibrium velocities in the initial jam,
the initial jam densities $\rho\sub{jam}\sup{(a/b)}=140$\,vehicles/h
and $\rho\sub{jam}\sup{(c/d)}=105$\,vehicles/h were chosen
differently.
In the viscosity traffic model , the finite velocity diffusion
leads to an increase of velocity also in the congested part (see
(d)) and to a subsequent further increase of density (see (c)). The
ensuing positive density gradient would lead to negative velocities
for an initial jam density higher than 106\,vehicles/km, which is
much lower than the maximum density $\rhomax=160$\,vehicles/km.}
\end{figure}

The main difference between the gas-kinetic-based traffic model and
other macroscopic traffic models is the nonlocal character of the
braking term. The nonlocal term in
Eq.~(\ref{Vedyn}) has smoothing properties similar to those of a
viscosity term, but its effect is forwardly directed and,
therefore, more realistic. In contrast, models with an explicit
viscosity term\cite{lee} lead to unphysical humps in the vehicle
density (see Figure~\ref{figModel1}) and even negative
velocities\cite{requiem} (see Figure~\ref{figModel2}).
Our model also has favorable properties with respect to
numerical stability and integration speed, and hence allows a
robust real time simulation of freeway stretches up to several
thousand kilometers on a personal computer.

\smallskip \noindent {\bf Some Explicit Finite Difference Methods}

A desirable property of the above traffic equations is that they can
be formulated in terms of a conservation equation 
with a sink/source term $\vec{s}$:
\begin{equation}
 \frac{\partial \vec{u}}{\partial t} + \frac{\partial \vec{f}(\vec{u})}
 {\partial x} = \vec{s}(\vec{u}) \, .
\end{equation}
This form makes it possible to use a variety of numerical standard
methods developed for the simulation of hydrodynamic
problems.\cite{Anderson,NumRec,Leveque} The
conservative form of the traffic equations reads
\begin{equation}
\label{rhocons}
 \frac{\partial \rho}{\partial t} + 
 \frac{\partial Q}{\partial x} 
 = \pm \nu^\pm(x,t),
\end{equation}
and
\begin{equation}
\label{Qcons}
\frac{\partial Q}{\partial t} 
 + \frac{\partial (Q^2/\rho + P)}{\partial x} = \frac{\rho V_{\rm e} - Q}{\tau}
 \pm \nu^\pm V^\pm \, , 
\end{equation}
where $Q(x,t) = \rho(x,t) V(x,t)$ is the traffic flow and
$V^\pm$ denotes the average velocity of vehicles
which enter (+) or leave ($-$) the freeway at ramps.
For the case $V^\pm \ne V$, we obtain
the additional terms $\pm \nu^\pm (V^\pm - V)/\rho$ in
(\ref{geschwin}). We have
\begin{eqnarray}
 \vec{u} &=& \big [\rho,Q\big ] \, , \\ 
 \vec{f} &=& \big [Q, (Q^2/\rho + P)\big ] \, , \\
 \vec{s} &=& \big [\!\pm \nu^\pm, (\rho V_{\rm e} - Q)/\tau 
 \pm \nu^\pm V^\pm\big ] \, .
\end{eqnarray}

For the explicit numerical solution methods we will discuss, $x$ and
$t$ are discretized with uniform values of
$\Delta x$ and $\Delta t$, respectively. Hence, we calculate 
$\vec{u}$ at the discrete points $(j\,\Delta x,n \, \Delta t)$
with $j,n \in \{0,1,2,\dots\}$. For brevity, we use the notation
$\vec{u}_j^n = \vec{u}(j \, \Delta x, 
n\,\Delta t)$. We will discuss the following numerical integration 
methods:\cite{Anderson,Leveque} \hfill \break
{\em 1. Lax-Friedrichs method}
\begin{equation}
\vec{u}_j^{n+1}
= \frac{\vec{u}_{j-1}^n +\vec{u}_{j+1}^n}{2}
 - \frac{\Delta t}{2\Delta x} (\vec{f}_{j+1}^n - \vec{f}_{j-1}^n) 
+ \Delta t \, \vec{s}_j^n \, . 
\end{equation}
{2. \em Upwind method}
\begin{equation}
\vec{u}_j^{n+1} = \vec{u}_j^n 
 - \frac{\Delta t}{\Delta x} (\vec{f}_{j}^n - \vec{f}_{j-1}^n)
 + \Delta t \, \vec{s}_j^n \, .
\end{equation}
{\em 3. MacCormack method}
\begin{eqnarray}
 \tilde{\vec{u}}_j^n &=& \vec{u}_j^n
 - \frac{\Delta t}{\Delta x} (\vec{f}_{j}^n - \vec{f}_{j-1}^n)
 + \Delta t \, \vec{s}_j^n \, , \hskip 4.7pc {\rm
(predictor)}
\\
\vec{u}_j^{n+1} &=& 
 \frac{1}{2} \bigg[ \tilde{\vec{u}}_j^n + \vec{u}_j^n
 - \frac{\Delta t}{\Delta x} (\tilde{\vec{f}}_{j+1}^n 
 - \tilde{\vec{f}}_{j}^n) + \Delta t \,
\tilde{\vec{s}}_j^n \bigg] \, . \quad {\rm (corrector)}
\end{eqnarray}
{\em 4. Lax-Wendroff method}
\begin{eqnarray}
 \vec{u}_{j+\frac{1}{2}}^{n+\frac{1}{2}}
 &=& \frac{1}{2} \bigg[ \vec{u}_j^n + \vec{u}_{j+1}^n 
 - \frac{\Delta t}{\Delta x} (\vec{f}_{j+1}^n - \vec{f}_j^{n})
+ \frac{\Delta t}{2}\,(\vec{s}_j^n + \vec{s}_{j+1}^n) \bigg] \, \qquad {\rm
(predictor)} \\
\vec{u}_j^{n+1}
 &=& \vec{u}_j^n - \frac{\Delta t}{\Delta x} 
 \Big [ \vec{f}_{j+\frac{1}{2}}^{n+\frac{1}{2}}
 - \vec{f}_{j-\frac{1}{2}}^{n+\frac{1}{2}} \Big ] + \frac{\Delta t
}{2}\, 
\Big [\vec{s}_{j+\frac{1}{2}}^{n+\frac{1}{2}}
 +\vec{s}_{j-\frac{1}{2}}^{n+\frac{1}{2}}\Big ] \, . \hskip 3.7pc
{\rm (corrector)}
\end{eqnarray}

\smallskip {\it Consistency Order}. If the spatial
variation of $\vec{u}$ is sufficiently smooth, the Lax-Friedrichs
and upwind methods are first-order, that is, the upper bound of the
local error is proportional to $\alpha$ if
$\Delta x$ and $\Delta t$ are simultaneously decreased by a factor
of $\alpha$. In general, the upwind method is not stable and the
differential operator needs to be decomposed into parts which are
treated by upwind and downwind differencing, respectively (the
Godunov method\cite{Leveque,Lebacque,Ans,BuNeNa}). Fortunately, for
traffic equations of the form \refkl{geschwin}, the upwind method
is stable (and equivalent to the Godunov method).

The two-step MacCormack and Lax-Wendroff methods are second-order,
that is, the upper bound of the local error is proportional to
$\alpha^2$. However, for shock-like solutions, the order is lower 
for all of the above integration methods.\cite{Leveque} 
Note that the predictor step
of the MacCormack method is an ordinary upwind step, and the
corrector step consists of the average between the predictor and a
``downwind'' step with
$\vec{f}$ calculated with the values of the predictor.
Interchanging the order of the upwind and downwind differencings of
the two steps has nearly no effect for the equations investigated
here.

\smallskip {\it Accuracy}. Although the discretization
errors associated with nonlinear equations are difficult to
determine, good estimates are usually obtained by doing a local
linearization, at least for smooth solutions.\cite{Leveque} 
Because realistic traffic models can
produce sharp gradients, but no real shock fronts, the
linearization is applicable to the numerical treatment of
macroscopic traffic equations. This analysis shows that the upwind
method is more accurate than the Lax-Friedrichs method.

\begin{figure}
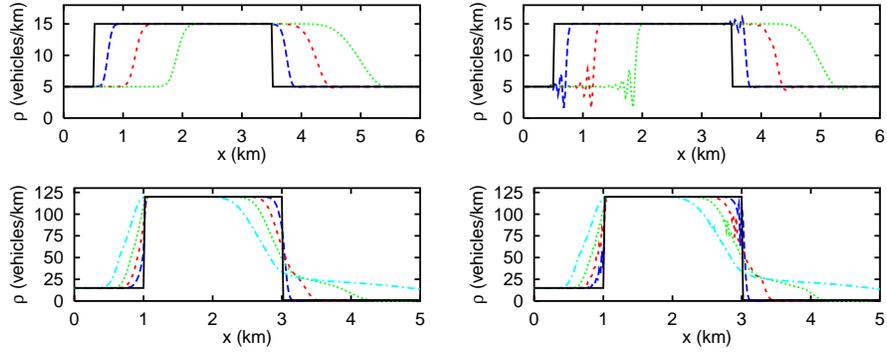

 \begin{center}
\hspace*{-8\unitlength}
\includegraphics[width=170\unitlength]{fig3a.cps}
\includegraphics[width=170\unitlength]{fig3b.cps}\\[-8\unitlength]
\hspace*{-8\unitlength}
\includegraphics[width=170\unitlength]{fig3c.cps}
\includegraphics[width=170\unitlength]{fig3d.cps}
\end{center}
\caption{\label{figMethod} 
Comparison of the upwind method, (a) and (c), with the MacCormack
method, (b) and (d), for simulations of the nonlocal,
gas-kinetic-based traffic model with discontinuous initial
conditions (---).  The plots correspond to transitions between
free traffic at two different vehicle densities (see (a) and (b)),
and to transitions to and from congested traffic with backwards
propagating fronts (see (c) and (d)). The density is shown at
$t=0$\,s (---), 10\,s (--~--), 30\,s (-~-~-), 60\,s ($\cdots$), and
120\,s (--~$\cdot$~--) (only in (c) and (d)). In the MacCormack
method simulations, the oscillations behind the large gradient
result from the dispersion error (see (b)), whereas the
oscillations around $x=3$\,km originate from nonlinear
instabilities (see (d)).} 

\end{figure}

Although the main discretization error of the
Lax-Friedrich and upwind methods yields a numerical diffusion, which
causes a smoothening of shock fronts (see
Figures~\ref{figMethod}a, and (b)), for the MacCormack and
Lax-Wendroff methods there is numerical dispersion, which is
associated with a slower propagation of waves with small amplitudes
and can lead to oscillations in the density and flow fields behind
(but not before) large gradients (cf.\ Figures~\ref{figMethod}c
and (d)). However, nonlinear instabilities (see below) also may
lead to oscillations. An example is the
oscillations at the downstream front of the large amplitude jam
of Figure~\ref{figMethod}d.

\smallskip {\it Numerical Stability}. An integration method is
numerically unstable if errors grow exponentially, which usually
leads to wildly oscillating density profiles with very short wave
lengths and eventually to overflow error. Even in
the quasi-linear case, the above explicit discretization methods
can lead to three types of instabilities which implies three
conditions for numerical stability:
\begin{enumerate}
\item {\em Convective instability}. Instabilities of the finite
difference method for the flux term lead to the
Courant-Friedrichs-Lewy condition\cite{Anderson,Leveque} which for our model
becomes
\begin{equation}
\label{CFL}
 \left| { \Delta t} \right| \le \frac{\Delta x}{V_0} \, ,
\end{equation}
 where $V_0$ is the maximum average velocity.
For example, for $V_0 = 40$\,{\rm m/s} (144\,{\rm km/h}) and a
spatial discretization of $\Delta x = 20\,{\rm m}$, we obtain
$\Delta t
\le 0.5\,{\rm s}$.

\item {\em Diffusion instability}. Models which contain an explicit
viscosity term $D \partial^2 V/\partial x^2$,
change the otherwise hyperbolic character of the
partial differential equations to a parabolic one. For numerical
stability, the additional diffusional Courant-Friedrichs-Lewy
condition
$D \, (\Delta x)^2/(2\,\Delta t) \le 1$ must be
fulfilled.\cite{Anderson,Leveque} 
This condition does not apply to simulations
of the nonlocal, gas-kinetic-based traffic model.

\item {\em Relaxational instability}. Due to the finite $\Delta t$,
instabilities can develop for spatially homogeneous density and
flow fields if 
$\Delta t$ is larger than one of the local relaxation times
$1/r_k$, where $r_k$ are the eigenvalues of the functional
matrix of the sink/source term $\vec{s}$. This condition puts
further restrictions on $\Delta t$, which depend on the maximum
vehicle density in the simulation. 
\end{enumerate}

\begin{figure}
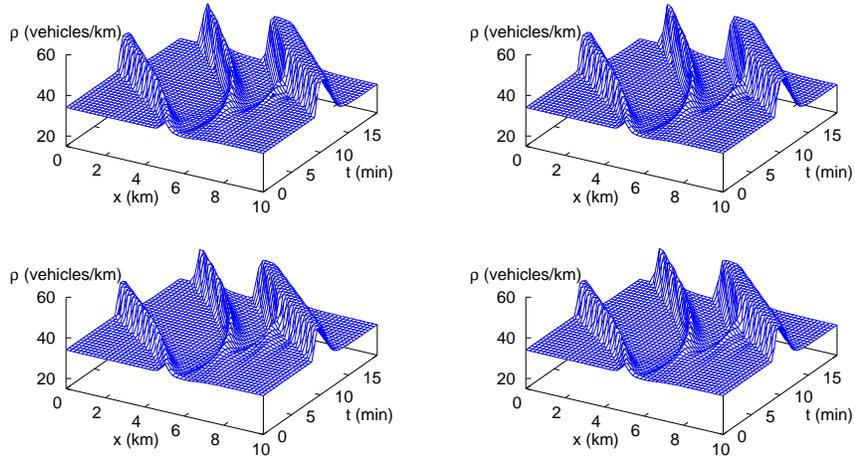

\begin{center}
\includegraphics[width=170\unitlength]{fig4a.cps}
\includegraphics[width=170\unitlength]{fig4b.cps}
\\[-20\unitlength]
\includegraphics[width=170\unitlength]{fig4c.cps}
\includegraphics[width=170\unitlength]{fig4d.cps}
\end{center}
\caption{\label{figDiscr} Spatio-temporal formation
of stop-and-go waves, simulated with the non-local, gas-kinetic-based 
traffic model,
using different integration methods and discretizations.
(a) MacCormack method with $\Delta t=0.4$\,s and $\Delta x=20$\,m.
(b) Upwind method with $\Delta t=0.4$\,s and $\Delta x=20$\,m.
(c) Upwind method with $\Delta t=0.01$\,s and $\Delta x=20$\,m. 
(d) Upwind method with $\Delta t=0.1$\,s and $\Delta x=5$\,m.  }

\end{figure}

Because viscosity and diffusion terms are replaced by a nonlocal term,
coarser discretization is possible in the above gas-kinetic-based 
traffic model than in most other
traffic models, allowing real time simulation of much larger systems.
For simulations of this model, good results 
are obtained for $\Delta x = 20$\,m and $\Delta t = 0.4$\,s
for the MacCormack as well as the upwind method.
Figure~\ref{figDiscr} shows that a finer discretization gives almost
identical results. 

The development of traffic instabilities starting with
almost homogeneous initial traffic 
is a very strict test of numerical accuracy (see
Figure~\ref{figDiscr}). In most situations, for example, when
simulating fronts or already developed congested traffic, the
accumulated discretization error is much smaller.

In addition to the quasi-linear instabilities discussed above,
genuine nonlinear instabilities may arise for certain numerical
methods and simulation conditions. An example is the oscillations
at the downstream front of the large amplitude jam of
Figure~\ref{figMethod}d. It turns out that second-order methods
are more sensitive to these instabilities. Note that, although the
quasi-linear behavior is the same for the MacCormack and
Lax-Wendroff methods, the nonlinear behavior can be
different.\cite{Anderson} For the above traffic equations, {\em the
MacCormack method is more stable than the Lax-Wendroff method}, and
hence we will use the MacCormack method whenever a second-order
method is desired.

When simulating the nonlocal model with one of the first-order
methods, we did not observe nonlinear instabilities. Furthermore, 
{\em higher order methods are not necessarily more
accurate},\cite{NumRec} especially for large gradients.
Thus, one should always implement different numerical
methods and compare their simulation results. Aside from their
double integration speed, it is sometimes preferable to use
first-order methods (causing somewhat smoothed wave fronts) 
instead of second-order methods (producing oscillations and
sometimes nonlinear numerical instabilities). One reason is
that oscillations act like perturbations, which may give rise to
additional traffic jams that do not correspond with reality.

\smallskip \noindent {\bf Initial and Boundary Conditions}

To calculate traffic flows, it is necessary to specify the
initial and boundary conditions. The
initial conditions are completely determined by specifying
$\v{u}(x,0)$ in a range $[0,L+\delta]$, where $L$ is the length of
the simulated section and $\delta$ is the maximum nonlocality,
which is of the order of 30\,m in the gas-kinetic-based model. In systems with
open rather than periodic boundaries, the initial conditions
usually influence the simulation results only for a short time, at
least, if we starts with free traffic, because errors in the
specification of the initial conditions are propagated outside the
simulated freeway stretch very quickly. Thus, the choice
of the initial conditions is unimportant. For example, we can
start with a linear interpolation of the measured initial boundary
values. However, because of conservation of vehicle
number, the initial conditions are relevant for closed systems with
{\em periodic boundary conditions}, which are given by
$\vec{u}(0,t) = \vec{u}(L,t)$, and $\partial
\vec{u}(0,t)/\partial x = \partial \vec{u}(L,t)/\partial
x$.

The specification of time-dependent boundary conditions is
much more involved. The following options are reasonable in
different situations: 
\begin{enumerate}
\item 
{\em Dirichlet boundary conditions} are given by the empirically
measured values $\vec{u}(0,t)$ and $\vec{u}(L,t)$
at both ends of the particular freeway stretch.  
For the values at $x\in [L,L+\delta]$ beyond the right
boundary (required by the nonlocal term of the gas-kinetic-based
model), a constant $\v{u}(x)=\v{u}(L)$ of the boundary
values is assumed.
\item
{\em Homogeneous von Neumann boundary conditions} assume that the
density and vehicular flow remain unchanged at the boundaries 
$x=0$ and $x=L$:
\begin{equation}
 \frac{\partial \vec{u}(0,t)}{\partial x} = 0 \, , 
 \quad \frac{\partial \vec{u}(L,t)}{\partial x} = 0 \, .
\end{equation}
Again, a constant value of $\vec{u}$ is assumed for $x>L$. 
\item
{\em Free boundary conditions}
assume that the traffic state is smooth at the boundary,
\begin{equation}
 \frac{\partial^2 \vec{u}(0,t)} {\partial x^2} 
 = \frac{\partial^2 \vec{u}(L,t)} {\partial x^2} = 0 \, ,
\end{equation}
with a linear extrapolation $\v{u}(L+\delta x) = \v{u}(L)
+ \delta x\, \partial\vec{u}(L)/\partial x$ for $\delta x\in[0,
\delta]$.
\item
{\em In- and outflows} $\pm Q_{\rm rmp}$
at on-ramps (+) and off-ramps ($-$),
can be considered as follows.\cite{Letter,GKT-scatter,Phase}
If $n$ is the number of freeway lanes (without ramps) and $L_{\rm
rmp}$ is the length of the ramp, we simply set
\begin{equation}
 \nu^\pm = \frac{Q_{\rm rmp}}{nL_{\rm rmp}} \, .
\end{equation}
\end{enumerate}

Generally speaking, periodic boundary conditions are best suited for
theoretical investigations of stability, and Dirichlet
boundary conditions are best for simulations of real traffic with
measured values of velocity and the traffic flow at the boundary.
Homogeneous von Neumann or free boundary conditions for
``absorbing'' boundaries should be used if 
the traffic situation outside the boundaries is not of interest. The
effects of the latter two boundary conditions are nearly the same.
An alternative to using open boundaries is to apply periodic
boundary conditions for distances which are much longer
than the freeway stretch of interest. 
However, for a simulated time interval
$T_{\rm sim}=2\,{\rm h}$, the required additional length is
order
$V_0 T_{\rm sim} \approx 200$\,km, which considerably reduces the 
efficiency of the numerical integration.

There are problems associated with using Dirichlet boundary
conditions. Imposing Dirichlet boundary conditions at both sides
usually leads to an overdetermined system, so that either numerical
instabilities occur or the boundary conditions are simply ignored
by the integration method. Even Dirichlet boundary conditions at
one of the two boundaries will lead to an unphysical, ill-posed
problem in certain situations. This case becomes clear when
imposing downstream Dirichlet boundary conditions for free traffic
flow at low densities. If the imposed boundary flow is higher than
the flow arriving from the simulated section at the boundary, the
continuity equation will lead to a decrease of the density, which
eventually results in unphysical negative densities. On the other
hand, if Dirichlet boundary conditions are imposed at the upstream
boundary in a situation of congested traffic where the boundary
flow is higher than the equilibrium flow, the continuity equation
leads to an increased density implying an even lower equilibrium
flow. This positive feedback will eventually lead to a divergence
of the density near the boundary.

We solve this problem by dynamically switching between Dirichlet
and von Neumann boundary conditions, depending on the density close
to the boundaries. The idea is to use Dirichlet boundary conditions
if the direction of information flow points towards the simulation
stretch, and von Neumann boundary conditions otherwise. Because the
traffic equations contain source terms like the relaxation term, 
the propagation direction of information cannot be obtained from
the characteristic velocities (which are the propagation
velocities of locally linearized homogeneous partial differential
equations like conservation equations without source terms.
Rather, the linearized group velocity
$v_{\rm g} = \partial Q_{\rm e}/\partial\rho$ of kinematic waves
gives a good criterion for separating the traffic situation into
free traffic ($v_{\rm g}>0$) whose information flow points
downstream, and congested traffic ($v_{\rm g}<0$) where the
information flow points upstream. Based on this idea, the following
hybrid boundary conditions yield good results: If the upstream
boundary values
$\rho_0(t)$ and $Q_0(t)$ satisfy
\begin{equation} \label{BChybrid}
\rho_0(t)\le\beta_1 \rho_{\rm m} \quad
 \mbox{or} \quad Q_0(t) < \beta_2 Q(\Delta x,t) 
\end{equation}
Dirichlet boundary
conditions are used; $\rho_{\rm m}$ is the density associated with
maximum equilibrium flow,
$\beta_1 = 0.95$, and $\beta_2 =0.98$. Otherwise, homogeneous
Von-Neumann boundary conditions are applied. For the downstream
boundary conditions, the directions of the inequalities are
exchanged. Note that these boundary conditions include situations
where Dirichlet boundary conditions are necessary at both sides (for
example, when a jam enters the downstream boundary of the
simulation region) as well as situations where no Dirichlet
boundary conditions are allowed at all (when congested traffic
formed at an inhomogeneity within the simulated region reaches the
upstream boundary). 

We illustrate the effects of different boundary conditions by a
simulation of the German motorway A8 near Munich with real traffic
data using the upwind method. Figure~\ref{figBC}a shows a
space-time plot of the density with Dirichlet
boundary conditions on both sides. The wing-like regions of higher
density represent ``synchronized'' 
congested traffic\cite{Kerner-sync} forming
upstream of a flow-reducing inhomogeneity,\cite{Phase} while the
region of the highest density represents a traffic jam entering
through the downstream boundary. Note that the unnecessary
downstream boundary conditions for
$t<$\,17:30\,h are ignored by the upwind differencing, whereas the
traffic jam entering the downstream boundary for
17:30\,h\,$<t<$\,18:20\,h is accepted. This desirable behavior
cannot be expected in general. For example, the MacCormack method is
unstable in this situation. 

Figure~\ref{figBC}b demonstrates the remarkable property that
the boundary conditions for the density are nearly irrelevant.
Although the boundary conditions for the flow are the same as in
Fig.~\ref{figBC}a, a constant amount of 20\,vehicles/km has been
added artificially to the upstream boundary conditions for the
density. Nevertheless, the traffic dynamics, in particular the
spontaneous breakdown at $x\approx 40$\,km and
$t\approx$\,17:00\,h, is nearly unchanged. This behavior can be
explained by recognizing that the inflow into the freeway and the
outflow from it determine the number of vehicles on the freeway
stretch, which cannot be changed by the traffic dynamics between
the boundaries. Hence, changes of the traffic flow are far reaching,
while the influence of density (or velocity) boundary conditions is
restricted to the distance that the vehicles drive during the
relaxation time. 

Figure~\ref{figBC}c illustrates the effect of using homogeneous
von Neumann downstream boundary conditions instead of Dirichlet
boundary conditions. For
$t<$\,18:00\,h, the situation is identical to
Figure~\ref{figBC}a. For
$t>$\,18:00\,h, however, the upstream moving traffic jam passing
the downstream boundary is ignored. Hence, the simulation result is
very different when the direction of the downstream information
flow changes. 

Finally, Fig.~\ref{figBC}d shows a simulation with the hybrid
boundary conditions (see Eq.~(\ref{BChybrid})) 
and the upstream boundary shifted
downstream by 3.5\,km. In this case, the traffic jam reaches the
upstream boundary, which is handled by the hybrid boundary
conditions, while all other boundary conditions would lead to a
numerical instability for any integration method.

\smallskip \noindent {\bf Summary}

We have seen that traffic flows are characterized by the occurrence
of congested regions, jams, and stop-and-go waves which are
associated with large gradients of the vehicle density and average
velocity. This rich behavior has stimulated an intense
research activity. We have discussed the problems of macroscopic
traffic simulations which are related to the numerical
integration of systems of coupled nonlinear partial differential
equations. Despite the differences with hydrodynamic equations, many
integration methods developed for conservation equations turn out
to be applicable. Compared to implicit methods,\cite{Anderson} 
explicit methods are less robust, but much more flexible with
regard to time-dependent boundary conditions and (variational)
optimization problems, and are usually
computationally faster. Among the explicit first-order methods, the
upwind method is more accurate than the Lax-Friedrichs method. The
MacCormack and the two-step Lax-Wendroff methods are second-order,
but they are less efficient by a factor of two and produce
unrealistic oscillations close to steep gradients. 
The numerical precision may be improved by ``high
resolution methods,''\cite{Leveque} where a first-order method is
used for large gradients and a second-order method is used for small
gradients. This approach may combine the accuracy of the
second-order methods with the smoothness of the first-order methods.

The most important factor that determines the
computation time is the choice of the traffic model. In particular,
the simulation of the nonlocal, gas-kinetic-based traffic 
model is significantly more efficient than the numerical solution of
models with viscosity or diffusion terms. This efficiency is mainly
related to the fact that the 
diffusional Courant-Friedrichs-Lewy condition, which does
not apply for nonlocal terms, is usually far more restrictive than
the other instability conditions, especially for fine
discretizations. Because diffusion terms also produce unrealistic
effects close to steep gradients, it may be reasonable to generally
replace models with diffusion or viscosity terms by
nonlocal models. Anyway, diffusion and viscosity terms
are often a lowest-order approximation of nonlocal terms. They are
mainly used for historical reasons, because they can be better
treated analytically. 

Finally, we have discussed suitable specifications of the
boundary conditions. 
In most previous simulation studies, periodic boundary conditions 
were used to circumvent the intricate problems related to open
boundaries, which are required for the simulation of real freeways.
We found that Dirichlet boundary conditions work in some cases, but
fail in others, because of overspecification. The most successful
treatment is based on hybrid boundary conditions which switch
between Dirichlet and homogeneous von Neumann (or free) boundary
conditions depending on the respective direction of information
flow.
%
%In conclusion, new concepts in the simulation of traffic flows
%allow the calculation of large traffic networks with more realistic
%traffic models and more efficient finite difference methods.
%We can now design on-line controls for
%efficient traffic optimization, to calculate the environmental
%impact of congestion, and to develop methods for traffic
%forecasts.

\smallskip \noindent {\bf Acknowledgments}

The authors are grateful for financial support by the DFG 
(Heisenberg scholarship He~2789/1-1) and by the BMBF (research
project SANDY, grant No.~13N7092). D.\ H.\ also thanks Tamas Vicsek
for his warm hospitality at the Institute of Biological Physics,
where part of this column was written.

\smallskip \noindent {\bf Suggestions for Further Study}

\smallskip

\noindent 1. {\it Parameter-dependence of flow-density relation}.
(a) Find the density dependence of the stationary, homogeneous
solution for the average velocity $V_{\rm e}$ in the nonlocal,
gas-kinetic-based traffic model by
setting the temporal and spatial derivatives to zero and solving
for $V$. (b) The flow-density relation $Q_{\rm e}(\rho) = \rho
V_{\rm e}(\rho)$ depend on which of the
model parameters, $V_0$, $\tau$, $\gamma$, $T$, and $\rho_{\rm
max}$? Which parameters are irrelevant for the stationary
homogeneous solution? (c) Plot
$Q_{\rm e}$ and the velocity-density relation
$V_{\rm e}$ for different parameter sets. Start with the values
used in this column. Check that a speed limit (that is, a decrease
in $V_0$) reduces the flow at small densities, but only by a
negligible amount for medium and high densities. In addition,
compare pure car traffic with $V_0 = 130$\,km/h, $T=1.2$\,s, and
$\rho_{\rm max} = 160$\,vehicles/km with traffic containing a
considerable fraction of trucks with, for example, $V_0 =
90$\,km/h, $T=3$\,s, and
$\rho_{\rm max} = 110$\,vehicles/km. In which density regimes do
the flow-density relations almost agree, and in which 
are they very different?\cite{GKT,GKT-scatter}

\smallskip \noindent 2. {\it Stability of homogeneous traffic with
respect to a localized perturbation}. Simulate freeway traffic for a
circular road of 10\,km circumference with the nonlocal,
gas-kinetic-based traffic model and the parameters used in this column. Assume
homogeneous equilibrium traffic of density
$\overline{\rho}$, and add to the density a localized perturbation 
of amplitude $\Delta \rho$ so that the initial conditions are given
by\cite{kk-94}
\begin{eqnarray*}
\rho(x,0) &=& \overline{\rho}
+ \Delta \rho \biggl[
\cosh^{-2} \biggl(\frac{x-x_0}{w^+} \biggr) 
- \frac{w^+}{w^-} 
 \cosh^{-2} \biggl(\frac{x-x_0-\Delta x_0}{w^-} \biggr) 
\biggr], \\ Q(x,0) &=& Q_{\rm e}(\overline{\rho}) \, .
\end{eqnarray*}
Here, $x_0$ and $x_0 + \Delta x_0$ are the positions of the
positive and negative peaks of the perturbation with widths
$w^+$ and
$w^-$, respectively. Choose $w^+ = 200$\,m, $w^-=800$\,m, and
$\Delta x_0 = w^+ + w^- = 1000$\,m. Use linear interpolation to
calculate the nonlocal terms and take into account that the
circular road implies that 
$\vec{u}(L+\delta x)=\vec{u}(\delta x)$. 
Perform the simulations with the upwind method on a fixed grid with
spatial and temporal discretizations of
$\Delta x=20$\,m and
$\Delta t = 0.4$\,s.

(a) Choose a small perturbation amplitude of $\Delta \rho=1$
vehicle/km and run the simulation for 30 minutes.
Check that the perturbation does not grow 
for $\overline{\rho}< \rho_{\rm c2}$ and
$\overline{\rho} > \rho_{\rm c3}$ with $\rho_{\rm c2} = 29$\,
vehicles/km and $\rho_{\rm c3} = 47$\,vehicles/km, but gives
rise to large-amplitude stop-and-go waves for
$\rho_{\rm c2} \le \overline{\rho} \le \rho_{\rm c3}$.

(b) Use larger perturbations of amplitudes up to $\Delta
\rho=60$ vehicles/km and show the traffic flow is metastable. To
do so, show that the model produces a single traffic jam for 
$\overline{\rho} \in (\rho_{\rm c1},\rho_{\rm c2})$ with
$\rho_{\rm c1}=27$\,vehicles/km and a dipole-like localized
structure for $\overline{\rho} \in (\rho_{\rm c3},
\rho_{\rm c4})$ with $\rho_{\rm c4}=50$\,vehicles/km, if $\Delta
\rho$ exceeds a (density-dependent) critical amplitude;
homogeneous traffic flow is found for subcritical perturbation
amplitudes.\cite{kk-94,GKT}

(c) In sufficiently large systems, there exists a subset of
densities
$\overline{\rho} \in (\rho_{\rm cv},\rho_{\rm c3})$ 
in the linearly unstable regime where traffic is convectively
stable.\cite{Phase} This stability means that the localized
perturbation will disappear for $t\to\infty$ at any given location
$x$ while, nevertheless, the global maximum of the perturbation
grows (because the system is linearly unstable). In our case,
$\rho_{\rm cv}$ is given by the value of
$\overline{\rho}$ at which the downstream boundary of the perturbed
zone does not move. Illustrate the case of convective stability by
making a three-dimensional plot of $\rho(x,t)$ of a simulation
running for 15 minutes with $\overline{\rho}= 45$\,vehicles/km and
$\Delta\rho=1$ vehicles/km. Show that $\rho_{\rm cv} =
42.5$\,vehicles/km.

\smallskip \noindent {\it 3. Characteristic parameters.\cite{GKT}}
Assume a circular road, model parameters, and initial conditions
as in Problem~2. 

(a) Evaluate the densities of fully developed traffic jams and of
free traffic at the outflows of jams. Determine the associated
flows and the propagation velocity of jam fronts. Hint: To obtain
the densities and flows, run the simulation for 60 minutes and
analyze the fields $\rho(x,t)$ and $Q(x,t)$. The jam
density (density of free traffic) is simply the global maximum
(minimum) of the density, and the jam flow (outflow from jams) is
the global minimum (maximum) of the flows. To obtain the group
velocity, save the density as a function of time at a given
position (for example,
$x=10$\,km) and determine the time intervals
needed for the jams to propagate around the circular
road. Use the last interval to avoid effects of transients.

(b) Plot the densities, the flows, and the propagation velocity in
separate plots as a function of the average density
$\overline{\rho}$ for a perturbation amplitude $\Delta \rho =
10$\,vehicle/km. What do you find? Check that there is a density
range in which the density inside and outside of traffic jams, the
associated flows, and the propagation velocity are independent of
$\overline{\rho}$, so that these quantities are characteristic
parameters of traffic flows. 

(c) Show numerically that the traffic inside and outside of jams is
nearly in equilibrium. Also show that the group velocity
can be expressed analytically in terms of the densities and flows
inside and outside of jams. This relation means that only two of the
characteristic parameters, for example, the group velocity and the
outflow from jams, are independent quantities.

(d) Do Problems 2a, 2b, and 3a for different values of the
relaxation time $\tau$. Show that with increasing $\tau$, the
region of linearly unstable traffic and the regions of metastable
traffic and the jam density increase, while the
outflow from jams decreases. In particular, homogeneous traffic is
always stable for
$\tau\le 18$\,s.

\smallskip \noindent {\it 4. ``Synchronized'' congested
traffic.\cite{Letter}} This traffic state occurs if a perturbation
reaches a stationary inhomogeneity of the road which can be an
on-ramp for
example.

(a) Simulate a freeway of length 10\, km with open
boundaries and an on-ramp of length $L_{\rm rmp}=400$\,m at $x_{\rm
rmp}=5$\,km by assuming in Eqs.~\refkl{rhocons} and \refkl{Qcons} a
source term $\nu^+ = Q_{\rm rmp}/L$ inside the ramp region
$x\in[x_{\rm rmp}-L/2, x_{\rm rmp}+L/2]$, but $\nu^+ = 0$ otherwise.
Assume $\tau=40\,{\rm s}$, the usual values for the other model
parameters, and homogeneous equilibrium traffic of density
$\overline{\rho}=15$ vehicles/km without perturbations as initial
conditions.
To obtain stationary conditions, simulate the first 20 
minutes with a constant ramp flow $Q_{\rm rmp} = 500$\,vehicles/h
using homogeneous von Neumann boundary conditions 
at both sides of the road.
Now, introduce a perturbation of amplitude $\Delta Q_{\rm
rmp}=150$\, vehicles/km by linearly increasing the ramp flow up to
$Q_{\rm rmp} + \Delta Q_{\rm rmp}$ at $t=22.5$\,min, and decreasing
it again to $Q_{\rm rmp}$ at $t=25$\,min. After running the
simulation for a total of 60 minutes, you should see ``synchronized''
congested traffic, that is, an increasing region of high density
and low velocity, but relatively high flow, whose upstream front is
propagating upstream, while the downstream front is pinned at the
ramp.

(b) Verify that ``synchronized'' congested traffic is in equilibrium and
determine the numerical value of its outflow $\tilde{Q}_{\rm out}$.
Show that the propagation velocity $v_{\rm g}$ of the upstream
front can be expressed by the relation
\begin{displaymath}
v_{\rm g} = \frac{\tilde{Q}_{\rm out} - Q_{\rm rmp} - Q_{\rm main}}
 {\rho_{\rm cong}(\tilde{Q}_{\rm out} - Q_{\rm rmp})
 -\rho_{\rm free}(Q_{\rm main}) } \, ,
\end{displaymath}
where $Q_{\rm rmp}$ and $Q_{\rm main}=Q_{\rm e}(\overline{\rho})$ 
are the inflows at the ramp and to the main road. The densities
$\rho_{\rm cong}(Q)$ and $\rho_{\rm free}(Q)$ denote, in accordance
with $Q_{\rm e} = \rho V_{\rm e}(\rho)$, congested traffic
($\rho \ge 31$\,vehicles/km) and free traffic ($\rho <
31$\,vehicles/km), respectively.

\smallskip \noindent {\it 5. Different traffic states close to an
on-ramp.\cite{Phase}} Varying the initial traffic density, the
length of the on-ramp, and the ramp flow leads to
several interesting states of congested traffic.
Do simulations as in Problem~4 with 
(a) $\overline{\rho}=20$ vehicles/km,
$Q_{\rm rmp} = 250$ vehicles/h, $\Delta Q_{\rm rmp}=150$ vehicles/km,
(b) $\overline{\rho}=20$ vehicles/km,
$Q_{\rm rmp} = 150$ vehicles/h, $\Delta Q_{\rm rmp}=150$ vehicles/km,
and (c) $\overline{\rho}=17$ vehicles/km,
$Q_{\rm rmp} = 250$ vehicles/h, $\Delta Q_{\rm rmp}=550$ vehicles/km.
You should observe (a) oscillating congested traffic, 
(b) triggered stop-and-go waves, and 
(c) pinned localized congestion. By varying the inflow
$Q_{\rm main}=Q_{\rm e}(\overline{\rho})$ to the main road,
verify that oscillating congested traffic emerges, if the
expression for
$v_{\rm g}$ in Problem~4 is negative (upstream moving front);
otherwise pinned localized congestion or free traffic occur.

\newpage

\newpage

%\clearpage
\newpage
%\clearpage

\newpage
\begin{figure}
\begin{center}
\includegraphics[width=170\unitlength]{fig5a.cps}
\includegraphics[width=170\unitlength]{fig5b.cps}
\\[-20\unitlength]
\includegraphics[width=170\unitlength]{fig5c.cps}
\includegraphics[width=170\unitlength]{fig5d.cps} 
\end{center}
\caption{\label{figBC} 
Simulation with empirical boundary conditions
at the German freeway A8 near Munich.
with the model parameters
$V_0 =130$ km/h for $x\le 41$ km,
$V_0 =97$ km/h for $x>41$ km,
$\tau=32$ s,
$T=2.0$ s,
$\rhomax=110$ vehicles/km,
and $\gamma=1.2$,
$A_0=0.008$, 
$\Delta A = 0.01$,
$\rho_{\rm c} = 0.27\rho_{\rm max}$,
and $\Delta\rho = 0.1\rho_{\rm max}$. Note that the decreased
desired velocity for $x>41$\,km reflects a 
flow-reducing gradient on the freeway stretch. 
(a) Dirichlet boundary conditions, applied to $\rho$ and $Q$ at
the upstream and downstream boundaries. We can distinguish two
different kinds of congestion: a region of ``synchronized''
traffic\protect\cite{Letter,Phase} with increased homogeneous
density, and a backwards propagating traffic jam entering
the freeway stretch at the downstream boundary. Dirichlet
boundary conditions are applicable here, because the traffic jam
does not reach the upstream boundary. (The jam dissolves before, as
the inflow decreases in the course of time.) (b) As in (a), but
with a constant
$\Delta\rho=20$\,vehicles/km added to the measured density at the
upstream boundary. The congested traffic enforced at the boundary
relaxes to free traffic very quickly. Note that the downstream
traffic patterns do not change significantly, because the
boundary flows (and, hence, the number of entering and leaving
vehicles per time unit) are the same as in (a). Nevertheless, this
result suggests that the congested traffic patterns are
self-organized structures. (c) As in (a), but with homogeneous von
Neumann boundary conditions for flow and density at the downstream
boundary, which ignore the upstream moving traffic jam entering at
the downstream boundary. However, the homogeneous congested region
of ``synchronized'' traffic is correctly reproduced, which indicates
that it is not triggered by the downstream boundary condition, but
it is rather generated by the inflow to the motorway at the upstream
boundary together with the flow-reducing gradient beginning at
$x=41.0$\,km. (d) Simulation with the hybrid boundary conditions
\protect\refkl{BChybrid}. Compared to (a), we shifted the upstream
boundary by 3.5\,km in downstream direction, so that it is reached
by the jam. The resulting spatio-temporal traffic patterns are
nearly identical as in (a), indicating that the hybrid boundary
conditions handle dynamically emerging congestion in a
natural way, even at the boundaries.}

\end{figure}

\end{document}